\documentclass{ws-ijmpd}
 
\begin{document}
 
\markboth{God{\l}owski}
{GLOBAL AND LOCAL EFFECTS OF ROTATION: OBSERVATIONAL ASPECTS}
 
%%%%%%%%%%%%%%%%%%%%% Publisher's Area please ignore %%%%%%%%%%%%%%%
%
\catchline{}{}{}{}{}
%
%%%%%%%%%%%%%%%%%%%%%%%%%%%%%%%%%%%%%%%%%%%%%%%%%%%%%%%%%%%%%%%%%%%%
 
\title{GLOBAL AND LOCAL EFFECTS OF ROTATION: OBSERVATIONAL ASPECTS}
 
\author{W{\l}odzimierz God{\l}owski}
 
\address{Institute of Physics, Opole University,\\
Oleska48, 45-052 Opole, Poland\\
godlowski@uni.opole.pl}

\maketitle
 
\begin{history}
\received{Day Month Year}
\revised{Day Month Year}
\comby{Managing Editor}
\end{history}
 
\begin{abstract}
In  the  paper we discussed the observational aspects of rotation
in  the  Universe on different scales. We show dependence between
the  angular  momentum  of the structures  and  their  size.  The
presented  observational situation is that  the  galaxies,  their
pairs  and  compact groups have a non-vanishing angular momentum.
In  the  structures of mass corresponding to groups of  galaxies,
this  feature  has  not been found, while  in  the  clusters  and
superclusters  alignment of galaxy orientation has been  actually
found.  Also we know that galaxies have net angular momentum
due to the fact that we actually measure the rotation curves of
galaxies. These facts lead to the conclusion that theories which
connect galaxy angular momentum with its surrounding structure
are at some extend favored by data. We  show
that  in the light of scenarios of galaxy formations this  result
could be interpreted as an effect of tidal forces mechanism,  but
it  is also consistent with Li's model, in which galaxies form in
the  rotating universe. The theoretical and observational aspects
of  possible  global rotation of the Universe were  discussed  as
well.
\end{abstract}
 

\section{Introduction}
 
    Investigating  occurrences of rotation in  the  Universe  and
its  possible  consequences  is  one  of  the  major  issues   in
contemporary astronomy. It involves the question of  rotation  of
individual cosmic structures as well as the question of  possible
rotation  of  the entire Universe. The rotating  objects  in  the
Universe  are found on various scales - from subatomic  particles
to  stars  and galaxies - so it is reasonable to ask whether  the
Universe  as a whole does rotate too. If there is the global,  or
just  a large-scale, rotation of the Universe, its effects should
be able to be detected in an observational way. In this work, the
possible  rotational effects at various astronomical  scales  are
discussed  and observational data that could be possibly  related
to rotation are presented.
 
    The  attempts  to observationally confirm that  the  Universe
actually  rotates  have been ever regarded much skeptically.  The
first  significant  evidence  for  the  global  rotation  of  the
Universe  was provided by Birch (1982). Birch considered position
angles  and  polarization of classic bright double radio  sources
and  found  that  the differences in position  angles  and  their
polarization are correlated with their position on the sky.  That
study  was immediately criticized by Phinney and Webster  (1983),
who  accused Birch of invalid application of statistical  methods
and  stated  that  his  data  are  inadequate  for  drawing  such
far-reaching  conclusions. In reply to them, Birch  (1983)  found
that  the effect observed by him had been present in the original
data  of  Phinney  and  Webster, while the  validity  of  Birch's
statistics was proved by Kendall and Young (1984). In  next  work
by  Phinney  et al. (1984), the data were reanalyzed using  novel
statistical methods and taking into account possible  errors  and
observational  uncertainties.  The  authors  concluded  that  the
effect  indicated by Birch was confirmed by observations, however
its nature was not clear. Bietenholz and Kronberg (1984) and
Bietenholz (1986) having extended the sample of analyzed  objects
did not find any evidence to confirm Birch's effect. Another attempt
to empirically  confirm the rotation of the Universe was undertaken
by Nodland and Ralston (1997a), who studied correlations  between
the  direction  and  distance to galaxies and the  angle  $\beta$
between  the  polarization direction and their  larger  axis  and
found  an  effect  which  they interpreted  as  rotation  of  the
polarization   plane  dependent  on  the  distance.   The   study
immediately  provoked discussion on the validity  of  statistical
methods  used  (Carrol  and  Field  1997,  Loredo  et  al.  1997,
Eisenstein and Bunn 1997, Nodland and Ralston 1997b, see also Ralston
and Jain 2004 for later review); it was also indicated  that  the
effect had not been confirmed by analysis of new, better observational
data (Wardle et al. 1997, see however that Jain and Ralston 1999
obtained opposite conclusion), and that even
if  Birch's  effect  were  to  be  regarded  as  evidence  for  a
large-scale rotation of the Universe, the value obtained by Birch
is  too  large  as  compared to the anisotropy  found  in  cosmic
microwave background radiation (CMBR). Potential difficulties  in
empirical confirmation of rotation of the Universe were mentioned
also  in a number of theoretical works. Silk (1970) already  drew
attention  to  the fact that at present, otherwise  than  in  the
early Universe, any dynamic effects of global rotation should  be
negligible,  and  the rotational period must  exceed  the  Hubble
time,  which  is a direct consequence of the small anisotropy  of
CMBR.  Later  on, Barrow, Juszkiewicz and Sonoda (1985)  obtained
strict   constraints   on  the  value  of  rotation   scalar   by
investigating  some  classes  of  Bianchi  models  which  involve
Friedmann' models as their special cases. They demonstrated  that
for  a flat universe there is a limit of $\omega/H_0\sim 2 \times
10^{-5}$ ($\omega\sim 1.5 \times 10^{-15} \, {\rm rad \, yr}^{-1}$). Recently
Pontzen   and  Challinor  (2007),  examining  effects   of   CMBR
polarization induced by global rotation, demonstrated  that  they
could  be  used to determine constraints on the amount of  global
rotation.  Su and Chu (2009) obtained a limit of this  kind  from
analyzing  the  2nd-order  Sachs-Wolfe effect,  namely  that  the
angular  velocity of shear-free rotation of $\Lambda CDM$ universe
is less than  $\omega  \sim 10^{-9} \, {\rm rad \, yr}^{-1}$ at the last
scattering surface. This constraint is weaker than that  obtained
by  Barrow,  Juszkiewicz and Sonoda (1985) for the  flat  Bianchi
models.  Chechin  (2010) investigated the rotational  effects  of
cosmic  vacuum. Considering the global rotation and  the  induced
rotation  of   elliptical  galaxies, he estimated  the  value  of
angular  velocity of the Universe in his model
as $\, \, \omega  \sim 10^{-19} \, {\rm rad \, s}^{-1}$
($3 \times 10^{-11} \, {\rm rad \, yr}^{-1}$). In  the
light  of  the above results the global rotation of the  Universe
cannot be regarded as confirmed by observations, thus instead  of
observational confirmation of rotation of the Universe one should
rather  talk  about rotational limits within a given cosmological
model.
 
    The  relativistic  models  with  rotation  and  geodetics  in
spacetime   behaving   according  to  general   relativity   were
investigated already in the first half of 20th century, beginning
from the works of Lanczos (1924), Gamow (1946) and Goedel (1949).
Ellis  and  Olive  (1983)  as  well as  Gron  and  Soleng  (1987)
demonstrated that the global rotation in inflationary  models  of
the  Universe  -  if  any  - should be small.  Demia{\'n}ski  and
Griszczuk  (1972)  discussed solutions of the Einstein  equations
for  the  case of flat homogeneous anisotropic space filled  with
expanding   and   rotating  ideal  fluid  with  non-zero   shear,
addressing  primarily  the question how rotation  influences  the
behaviour  of  matter near the initial singularity. The  Einstein
equations  with rotating fluid were also analyzed in a number  of
studies  by  Krasi{\'n}ski (Krasi{\'n}ski 1975, 1997,  1998a,b,c,
1999, 2001a,b). Fil'chenkov (2003, 2005) considered a possibility
of cosmological origin of rotation of astronomical objects due to
a   tunneling  effect  in  quantum  cosmology.  The  dynamics  of
Friedmann-Robertson-Walker (FRW) models with global rotation  was
discussed in Szyd{\l}owski and God{\l}owski (2005). The  role  of
rotation  of  objects  in the Universe and its  significance  for
astronomical  measurements was analyzed  in  Vishvakarma  (2006).
Recently  Bratek, Jalocha and Kutschera (2007) found a  class  of
explicit  solutions with cylindrical symmetry for  differentially
rotating gas. Discussion of anisotropic cosmological models  with
magnetic  field  and rotation can be found in  Demia{\'n}ski  and
Doroshkevich  (2007).Jain, Modgil, and Ralston (2007) examined the
Type 1a supernova data in order to determine if it shows any signal
of large scale anisotropy. The anisotropy was modeled by an extended
G\"{o}del metric (G\"{o}del-Obukhow metric), which incorporated
expansion along with rotation. Sousa, Pereira and Silva (2008)
considered the relation of energy  and  momentum density in Goedel's
universes, while Akarsu and Kilinc (2010) analyzed locally rotationally
symmetric (LRS) Bianchi Type I cosmological models are examined in
the presence of dynamically anisotropic dark energy and perfect fluid.
Iorio (2010) provided a  detailed account of influence of large-scale
 rotation on the Solar System.
 
    The   main  problem  in  testing  models  with  non-vanishing
rotation  is  that  there are no accepted  observables  for  this
purpose  to  date. It is also worth noting that while there  have
been established some upper limits for rotation from analysis  of
CMBR  and  primeval nucleosynthesis (Goedel 1949,  Hawking  1969,
Collins  and Hawking 1973, Barrow, Juszkiewicz, and Sonoda  1985,
Ciufolini and Wheeler 1995, Bunn, Ferreira and Silk 1996,  Kogut,
Hinshaw  and  Banday  1997), all these works  based  on  a  model
involving   both  shear  and  rotation,  which  makes  discussing
rotation a rather complicated task due to intertwined effects  of
shear  and  rotation.  Thus it is easier  to  study  a  Newtonian
equivalent  of  this  model, which allows for  rotation  with  no
shear.

\section{Cosmological models with dark radiation}
 
     In looking for limits of amount of rotation it is convenient
to  use Senovilla's formulation of Newtonian cosmology. According
to  Senovilla, Sopuerta and Szekeres (1998), density and pressure
in  a  Newtonian universe filled with homogeneous fluid will  not
explicitly depend on spatial variables, while depending on  time.
It  is,  however, assumed that fluid velocity depends in a linear
manner on spatial variables (Szekeres and Rankin 1977, Senovilla,
Sopuerta  and  Szekeres  1998). In this  case,  contrary  to  the
general  relativity solutions, there are solutions without  shear
that fulfill the Heckmann-Schucking equations with expansion  and
rotation. This situation has no equivalent in general relativity,
where   homogeneous   rotating  and  expanding   universes   (but
non-tilted - the models where vectors of 4-velocity $u^{\mu}$ are
not  orthogonal to the surface of homogeneity are called "tilted"
ones  -  Obukhov  (2000), Obukhov, Chrobok, and Scherfner  2002))
with  ideal  fluid have to possess a non-vanishing  shear  (Ellis
1966, King and Ellis 1973, Raychaudhuri 1979, Collins 1985). In a
Newtonian  universe  one can readily derive observables  required
for   cosmological  tests.  On  the  other  hand,  the  Newtonian
approximation seems a reasonable one, as it is revealed from  the
CMBR  analysis that the Universe is flat or nearly flat  (Spergel
et al. 2007), and furthermore, for the flat Bianchi I models, for
one,  the  shear  scalar decreases more rapidly (with  increasing
scale  factor) than the rotation scalar in an expanding  universe
with  ideal fluid as the field source (Hawking 1969, Ellis  1973,
Rothman and Ellis 1986, Li 1998).
 
     The motion of fluid in a homogeneous Newtonian universe  can
be  described  by the expansion scalar $\theta$, rotation  tensor
$\omega_{a,b}$,  and  shear  tensor  $\sigma_{a,b}$.  The  global
rotation  of  the  Universe consists here  in  homogeneous  fluid
rotating   as  a  whole  (Li  1998).  The  equation  representing
Newtonian  cosmologies, with expansion and rotation  but  without
shear, which should be regarded as the principal equation of  the
Newtonian  homogeneous cosmology is called the Heckmann-Schucking
equation  (Heckmann and Schucking 1959, Heckmann  1961).  It  was
Heckmann (1961) who demonstrated that such a Newtonian model  can
be constructed in a consistent manner.
 
    From   observations  of  distant  supernovae   of   Ia   type
(Perlmutter  et  al. 1998, 1999, Riess et al. 1998)  as  well  as
measurements of anisotropy of cosmic background (Spergel  et  al.
2006) it follows that the Universe accelerates its expansion  and
is  of  nearly  flat  geometry. It  raises  a  question  if  this
acceleration  is  due to rotating fluid and  if  the  results  of
Ia-type  supernovae observations can be explained by the  effects
of  rotation. Distance measurements, such as measuring luminosity
distance  as  a  function of redshift z,  are  sensitive  to  the
spatial geometry of the model and evolution of scale factor. Thus
the observed luminosity of a supernova (after taking into account
extinction,  K-term correction, etc.) will depend on the  current
density  of  various components of matter and their equations  of
state.  The  Heckmann-Schucking equation in  Newtonian  cosmology
plays   an   analogical  role  as  the  Friedmann   equation   in
relativistic  cosmology. From our point of view it  is  essential
that  in the Newtonian analogy of Friedmann equation the rotation
effect  produces  a  negative term scaling  as  $(1+z)^4$,  which
formally  corresponds  to  a presence of  (fictitious)  fluid  of
negative  density  scaling as radiation  and  thus  called  "dark
radiation". The presence of such a fluid would be much desirable,
since,  as it was demonstrated by Ichiki et al. (2002), it  could
contribute  to  accounting  for  the  discrepancies  between  the
abundances  of helium-4 and deuterium theoretically predicted  in
the  model  of primordial nucleosynthesis (BBN) and the  observed
ones.  A  preliminary  analysis (God{\l}owski  and  Szyd{\l}owski
2003b)  showed  that  while the rotation  effects  underlied  the
acceleration  of the Universe, they alone are not  sufficient  to
explain  the  observations  of Ia-type  supernovae,  which  would
require  additionally some kind of "dark energy",  for  instance,
the   cosmological  constant.  It  should  be  noted   that   the
constraints   on  rotation  in  God{\l}owski  and   Szyd{\l}owski
(2003b),  which could be derived from supernovae observations  at
that time, were quite weak.
 
    The   problem  of  determining  limits  on  rotation  of  the
Universe was addressed in a more detailed way in the next article
(God{\l}owski and Szyd{\l}owski 2006). In analyzing observational
effects of rotation, the Heckmann-Schucking equation was used  as
the  simplest  model in which they are separated from  the  other
effects and thus the influence of rotation can be considered in a
relatively easy way. This work took up observational limits to  a
negative term of $(1+z)^4$ type in the Friedmann equation (or its
Newtonian  equivalent  -  the Heckmann-Schucking  equation).  The
estimation  of  amount of rotation from the microwave  background
radiation  (CMBR) and primordial nucleosyntheis  (BBN)  was  made
assuming  the  presence  of additional, non-interacting  fluid  -
radiation  matter   (of  density parameter  $\Omega_{r,0}$).  The
constant   appearing  in  the  Heckmann-Schucking  equation   was
regarded  as  a  curvature  constant,  which,  as  the  Newtonian
cosmology  is  flat by definition and its curvature  constant  is
zero, was actually a departure from the Newtonian model.
 
    However,   in  that work it was noted that  a  negative  term
scaling as $(1+z)^4$ can be present in the Friedmann equation (or
its  Newtonian  equivalent) for a number  of  reasons.  The  main
interpretations here, apart from a cosmological model with global
rotation,   are  "dark  radiation"  in  Randall-Syndrum's   brane
scenario (Randal and Sundrum 1999, Vishwakarma and Sing 2003) and
the  Casimir  effect. Unfortunately, on the base of  astronomical
observations  only, it is not possible to distinguish  particular
components  of such a fluid scaling as radiation, as  cosmography
tests  the  Universe's  properties through the  relation  $H(z)$,
which does not allow for differentiating components if they scale
identically  with changing of the scale factor. Thus the analysis
yields  a  sum  of density parameters of such fluids  instead  of
constraints for pure rotation. So in the work of God{\l}owski and
Szyd{\l}owski  (2006) the $\Sigma \Omega_{dr,0}+\Omega_{r,0}$  of
density parameters of all (real and fictitious) fluids scaling as
radiation was estimated instead. It means that the limit  on  the
net contribution of density parameters from all such fluids, both
of  positive  (as  radiation) and negative (as Casimir's  effect)
energy  density. In the analysis, the observations of  supernovae
(Riess  et  al.  2004, Astier et al. 2005) as well  as  those  of
IIb-type Fanaroff-Riley radio galaxies (Daly and Djorgovski 2004)
were  used. Combining data from supernovae and radio galaxies  to
determine values of cosmological parameters was proposed by  Daly
and  Djorgovski  (2003), who also noted the  fact  that  for  the
purpose  of  investigations of this kind,   one  had  better  use
coordinate   distance  (Weinberg  1972)  instead  of   luminosity
distance,  commonly  used  in  analysing  data  from  supernovae.
Coordinate distance, as opposed to luminosity distance  does  not
depend  on  the value of Hubble constant $H_0$, but only  on  the
density parameters appearing in the model of (real or fictitious)
fluids. Using the coordinate distance in the work of God{\l}owski
and  Szyd{\l}owski (2006) had also this advantage that it allowed
for  making  a  consistent use of the constraints on cosmological
parameters  derived from the baryon oscillation peak  (Eisenstein
et  al.  2005) and the shift parameter in CMBR (Wang and  Tegmark
2004), which do not depend on the value of $H_0$ either.
 
    However,  while in the case of Fanaroff-Riley IIb-type  radio
galaxies Daly and Djorgovski provide coordinate distance for each
galaxy, for supernovae both Riess et al. (2004) and Astier et al.
(2005)  give  just  the  value  of  luminosity  module  (apparent
luminosity minus absolute luminosity), from which one can readily
obtain   luminosity  distance.  Still,  in  order  to   calculate
coordinate  distance  out of luminosity distance  one  needs  the
value  of  Hubble  constant.  In God{\l}owski  and  Szyd{\l}owski
(2006)  two  samples  of  supernovae  were  analyzed,  for  which
luminosity modules had been derived in a somewhat different  way.
For  Astier's  sample,  they  were  obtained  assuming  a  priori
$H_0=70 \, km \,s^{-1} \, Mpc^{-1}$ (Astier et al. 2005), and this
value was used in calculating  coordinate distances for supernovae
in this  sample, while  for Riess's sample the value of Hubble
constant should  be calculated  independently  beforehand (in  the
considered  model $H_0=64.4 \, km \,s^{-1} \, Mpc^{-1}$, or $h=0.644$
was obtained) and then used  in further calculations.
 
     Eventually, it was shown that the flat model of the Universe
with   density  parameter  $\Omega_{m,0}=0.3$  is   favoured   by
observations, while the value of $\Omega_{dr,0}$ parameter is  at
best  estimated as zero, which corresponds to vanishing rotation.
Thus the sought-after net negative contribution of fluids scaling
as  radiation, if any, is small. One can also obtain a  limit  on
the   value  of  density  parameter  $\Omega_{dr,0}$,  which   is
$\Omega_{dr,0}>-0.00035$ at the confidence level $0.95$, but this
means  that  the  sum  of density parameter  $\Omega_{dr,0}$  and
radiation  density  $\Omega_{r,0}$ ($\Omega_{r,0}  \sim  0.0001$,
Vishwakarma and Singh (2003)) is also negative. In that  case  we
would have the bouncing scenario instead of the Big Bang scenario
(Molina-Paris   and   Visser  1999,   Tippet   and   Lake   2004,
Szyd{\l}owski et al. 2005), where the Universe does not start its
evolution  from  the Bing Bang, but initially  contracts  to  its
minimal  size  (when the squared Hubble constants is  zero),  and
then  begins  to expand. It should be noted that  for  the  value
$\Omega_{dr,0}=-0.00035$ the Universe would bounce already at the
redshift z of about 1200, so the primordial nucleosynthesis would
not  have a chance to occur. Since the primordial nucleosynthesis
is  a  well-established aspect of cosmology,  it  imposes  strict
constraints  on  the  values of $\Omega_{dr,0}$.  Basing  on  the
estimates of Ichiki et al. (2003) it can be asserted  that at the
confidence   level  of  $0.95$,  the  primordial  nucleosynthesis
permits   the  values  $-7.22  \times  10^{-5}<\Omega_{dr,0}<0.65
\times  10^{-5}$.  So  there's a need of determining  a  stronger
limit   on  the  estimated  parameter  $\Omega_{dr,0}$.  Such   a
possibility is provided by analysing the first peak in the  power
spectrum of CMBR.
 
    It  should  be  stressed that there is a discrepancy  between
the   value   of  matter  density  parameter  $\Omega_{m,0}$   in
God{\l}owski and Szyd{\l}owski (2006), consistent with the  value
obtained earlier from supernovae observations (Riess et al. 1998,
Perlmutter  et  al.  1999,  Riess et al.  2005),  and  the  value
obtained by Spergel et al. (2006) from observations of background
radiation   ($\Omega_{m,0}=0.24$).  Furthermore,   there   is   a
discrepancy between the values of Hubble constant used  in  early
studies of the Universe's  acceleration based on observations  of
Ia-type  supernovae (Perlmutter et al. 1999,  Efstathiou  et  al.
1999,  Vishvakarma 2001) and the values $H_0$ obtained  from  the
CMBR analysis (Spergiel et al. 2007). It is usually accounted for
as  an  effect of uncertainty in determining the zero  point  for
Cepheids  used in calibrating the Hubble constant. For  instance,
Freedman  et al. (2001) suggest an inaccuracy of $\pm 8 km/s$  as
an  effect  of systematic errors. However, other researchers,  as
Tuner  (2010a,  2010b),  do  not  confirm  that  there  are  such
systematic errors involved.
    By  analyzing CMBR, Spergel et al. (2007) determined that the
density      parameter      $\Omega_{m,0}=0.128 \, h^{-2}$.   For
$\Omega_{m,0}=0.30$ (Perlmutter et al. 1998, 1999, Riess  et  al.
1998,  2004,  2007) the value of the Hubble constant  is  $H_0=65
\, km \,s^{-1} \, Mpc^{-1}$, but then for the flat Universe in the
classic scenario of  dark cold matter with a cosmological constant
($\Lambda CDM$) there  is a discrepancy between the theoretical
location of the first peak and the observed one.  The proper
location of the first peak   in  this  model  is  obtained for
 $H_0=73 \, km \,s^{-1} \, Mpc^{-1}$ ($\Omega_{m,0}=0.24$).
In the considered work of God{\l}owski and Szyd{\l}owski (2006) it
was shown that the presence of fluid  of negative density scaling
as radiation permits to avoid these discrepancies, and so we can
get a flat model  of  the  Universe with
$\Omega_{m,0}=0.3$ ($H_0=65 \, km \,s^{-1} \, Mpc^{-1}$), consistent with
both the Ia-type  supernovae  and CMBR observations. In this model
God{\l}owski and Szyd{\l}owski (2006) obtained at the  confidence
level  0.95  a strict constraint on the sum of density parameters
$\Omega_{dr,0}$  of negative fluids scaling as radiation:  $-1.05
\times 10^{-5}<\Omega_{dr,0}<-0.5 \times 10^{-5}$. This limit  is
stronger  than  that  obtained earlier  by  Ichiki  et  al.  from
analysis  of nucleosynthesis and CMBR. It should be also stressed
that  the  obtained  net  value  $\Omega_{dr,0}+\Omega_{r,0}$  is
greater than zero, which means that the Big Bang scenario and not
the bouncing one is valid indeed.
 
\section{Li's model and its predictions}
 
     In  1998 Li Li-Xin proposed a model involving galaxy forming
in  a  rotating universe. The very idea that as a consequence  of
the  conservation of angular momentum in a rotating universe, the
galaxies acquire its angular momentum during their formation, was
already  considered by Gamow (1946), Goedel (1949), and later  by
Colins  and  Hawking  (1973). One of its drawbacks  was  that  it
predicted alignment of galaxies' rotational axes, which  was  not
confirmed  observationally at that time. Li (1998)  explored  the
issue  of  galaxy  formation in a rotating  universe  in  a  more
detailed  manner. He analysed the model of homogeneous,  rotating
and   expanding   universe,  filled  with  ideal   fluid,   which
additionally  complies  with  the  laws  of  energy  and  angular
momentum  conservation. In Li's model, when the fluid's  equation
of  state  is $p=0$ (dust), the shear scalar $\sigma^2$ decreases
with increase of the scale factor as $a^{-6}$ (i.e. $\sigma^2 \sim
a^{-6}$),  while  the rotation scalar $\omega^2$ diminishes  more
slowly  ($\omega^2 \sim a^{-4}$). Thus it is quite reasonable  to
presume  that  (for a sufficiently large scale  factor)  that  in
comparison with rotation, shear is negligible, since it decreases
more  rapidly than the rotation scalar with increasing the  scale
factor.  The  model  is  based  on the  Steigman-Turner  equation
(Steigman and Turner 1983). Interesting comments on this equation
and its applicability can be found in Rothman and Ellis (1986).
 
     The relation between the mass of a cosmic structure and  its
angular  momentum is usually expressed by the empirical  relation
$J\sim  M^{5/3}$ (Wesson 1979, 1983, Carrasco, Roth  and  Serrano
1982,  Brosche 1986), which origin has been discussed for a  long
time.  One  of the first attempts to account it for was  made  by
Muradyan  (1975, 1980), who tried to explain it in terms  of  the
Ambarzumian's superdense cosmogony. Wesson (1981) pointed out its
possible  role  in  the unification of gravitation  and  particle
physics  and  later (Wesson 1983) argued that it is a consequence
of  self-similarity  of  Newtonian problem  applied  to  rotating
gravitationally  bound systems, while Mackrossan (1987)  involved
thermodynamical consideration for its explanation. Sistero (1983)
introduced  the  problem of rotational velocity of  the  Universe
into  considerations.  The same line  of  thought  was  taken  by
Corrasco  et al. (1982), who viewed the relation $J\sim  M^{5/3}$
as   a   consequence  of  mechanical  equilibrium   between   the
gravitational and rotational energies.
    Li (1998)  indicated  that in the model of rotating universe
the relation $J\sim  M^{5/3}$ is a direct consequence of the law of
angular momentum conservation. The correlation between the mass of
spiral galaxies  and their angular momentum was considered in a
detailed manner by a number of authors (e.g. Nordsiek 1973,
Dai, Liu and Hu 1978, Cerrasco, Roth and Serrano 1982, Abramyan and
Sedrakyan 1985). Investigating this correlation, Li (1998) estimated
the Universe's rotation and found the amount close to that  obtained
earlier by Birch (1982). Heavens and Peacock (1988) show that the
relation $J\sim  M^{5/3}$  is also a prediction of the tidal torque
scenario (such attempts were made, among  others, also by
Catelan and Theuns 1996).
 
    In  God{\l}owski et al. (2003a) it was pointed  out  that  in
Li's  model the relation between the angular momentum of a cosmic
structure  and  its mass is of a more complicated character  than
simply $J\sim M^{5/3}$ (which is just its approximation for large
masses  M). Actually the relation is of the form $J=k\,M^{5/3}-l\,M$,
where  k  and  l  are constants dependent on the  amount  of  the
Universe's  rotation, density of matter, size of a protostructure
and  the  moment  of  its  formation, as well  as  the  parameter
$\beta$,   determined  by  the  mass  distribution   within   the
protostructure. Consequently, the relation $J(M)$ has the  global
minimum $J_{min}$ for the mass $M_{min}$. It was also shown  that
$M_{min}$  does  not  depend on the amount  of  rotation  of  the
Universe,  while depending on the size of forming  protostructure
and the moment of  its formation (redshift $z_f$), as well as  on
the value of parameter $\beta$ (it is presumed to be close to 1).
Dark  matter  was  ignored in the considerations,  since  it  was
assumed   to  be  collisionless  and  non-interacting  with   the
observable matter  in any other way than by gravitational force.
Thus the mass estimates provided below pertain
to  the mass of dust structures (and not the total mass including
dark matter present in galaxies and clusters of galaxies).
 
    In  the  discussed work it was also demonstrated  that  under
reasonable  assumptions, the structures of certain masses  should
possess a vanishing angular momentum. One can then check  if  the
results  from  observations of galaxy structures  are  consistent
with  the predictions of Li's  model. But one should have in mind
numerous difficulties in theoretical estimating of $M_{min}$.  It
involves  a  number  of  approximations  since  values  of   such
parameters as mass distribution within the protostructures, their
size, and the moment of forming of galaxies and clusters are  not
known.  The  latter parameter depends strongly  on  the  accepted
scenario  of  formation of galaxies and their  clusters.  However
assuming some reasonable values of those parameters, one can  get
some  estimate of $M_{min}$. This theoretical minimum  should  be
then confronted with the relevant observational data.
 
    From  the  observational point of view  of  interest  is  the
absolute  value of angular momentum $|J|$, since in  general  the
direction of structures is not known, while one can quite readily
identify the situation when angular momentum vanishes or is  very
small.  The vanishing angular momentum J=0 is obtained  when  the
structure's mass is $M_0=2.15 \times M_{min}$.
    We  should  keep in mind that Li's model remains  valid  only
provided rotation occurs on a sufficiently large scale. Moreover,
due  to  its  scaling properties Li's model  can  be  applied  to
rotating  protostructures  of  various  initial  size.  Thus  our
considerations  concerning angular momenta of  galaxy  structures
will  be  also  valid  in  the  case  of  large-scale,  but   not
necessarily global, rotation of the Universe. So the model can be
of  interest even if we reject the hypothesis of global  rotation
of the Universe.
 
    In  God{\l}owski et al. (2003a) it was attempted to  estimate
the  masses of structures of minimal angular momentum in relation
to  the  size  of  protostructure. And thus, if  we  assume  that
clusters of galaxies arise as a result of collapse protostructure
(Sunyaev  and  Zeldovich  1972,   Dorohkevich  1973),  then   the
structures  of minimal angular momentum should be of  a  mass  of
galaxy  clusters.  In the case of protostructure  of  protogalaxy
size,  its mass should be of an order of globular cluster,  while
in  the  case of protosolar cloud it should be comparable to  the
giant moons of the Solar System. Except for the Moon, the angular
momenta  of such celestial object are actually smaller  than  the
angular momenta of either planets or asteroids (Wesson 1979).
 On should note however that formation of the stars inside the
galaxy or the formation of a solar system is such complex astrophysical
process that extrapolation of Li's (1998) and God{\l}owski (2003a)
results to a protosolar cloud is questionable.
 
    It   raises  a  question  if  there  is  any  possibility  to
observationally  test  the predictions of  Li's   model.  Such  a
possibility is provided by studies of galaxy plane orientation.

\section{Galaxy orientation and galaxy formation scenarios}
 
     Studies  of galaxy plane orientation were conducted  already
in   19th  century  (Abbe  1875).  The  methods  and  results  of
investigations performed up to Second World War can be  found  in
the article of Danver (1942), today only of historical value. The
first postwar work was the cited to this day treatise of Holmberg
(1946),  who  compared the numbers of galaxies seen  face-on  and
edge-on,  discussed the observational effects related to  optical
measurements of size of galaxy axes, and proved that the observed
excess  of  edge-on  galaxies is just  of  observational  origin.
Typical  for the early period after the Second World War are  the
analyses  of  Wyatt  and  Brown (1955) and  Brown  (1964,  1968),
devoted to investigating distributions of position angles  within
the galaxy-rich regions of the sky: Cetus (Wyatt and Brown 1955),
Pisces (Brown 1964), and in Hydra, Sextant, Ursa Maior, Virgo and
Eridanus  (Brown  1968). In his two studies  Brown  (1964,  1968)
discovered   a   departure  from  isotropy  of   position   angle
distribution.  Reinhardt (1970) and Reinhardt and Roberts  (1972)
analysed  distributions of position angles of large  semiaxes  of
the  galaxies from the Reference Catalogue of Bright Galaxies (de
Vaucouleurs  1964)  and found a very weak  preference  of  galaxy
plane  alignment with the equator plane of the Local Supergalaxy.
Nilson  (1974)  investigated the galaxies brighter than  $14.5^m$
from  his UGC catalogue (Nilson 1973). His results are consistent
with  those  obtained  by Reinhardt and Roberts.  However  either
results are compromised by the presence of galaxies not belonging
to the Local Supercluster in the analysed sample.
 
     Further important progress in investigations of galaxy plane
orientation  was  made by Hawley and Peebles  (1975),  who  in  a
detailed  manner  discussed the method  of  investigating  galaxy
orientation through analysing distribution of position angles  as
well  as possible errors and observational effects, in particular
the  earlier  results  of  Brown, indicating  their  insufficient
certainty due to possible observation errors.
 
     Investigating orientation of galaxy planes in  space  is  of
importance,  as various scenarios of formation and  evolution  of
cosmic  structures  predict different  distributions  of  angular
momenta   of   galaxies,   i.e.  provide  different   predictions
concerning  orientation  of  objects  at  different   levels   of
structure - in particular clusters and superclusters of galaxies.
This provides a method for testing scenarios of galaxy formation.
One  assumes that normals to the galaxies' planes are their  axes
of rotation, which seems to be quite reasonable, at least for the
spiral galaxies.
 
     In  the  scenario  of hierarchic clustering  (Peebles  1969,
Doroshkevich  1970,  Dekel 1985) large-scale  structures  in  the
Universe form "from bottom up", as a consequence of gravitational
interactions between galaxies. This means that galaxies  form  in
the  first  place and only later group into larger clusters.  The
galaxies'   spin  angular  momentum  arises  as  an   effect   of
interaction  with their neighbours. In original version of this
model the orientation of galaxy rotational axes would be random.
 One should note however that in the hierarchical clustering
theory naturally arises the Tidal Torque scenario. In this scenario
galaxies have their angular momentum perpendicular to the surrounding
structure due to the coupling between the protogalaxy region and
surrounding structure.
 
    On  the  other  hand, the scenario of primordial  turbulences
(von  Weizsacker 1951, Gamow 1952, developed further by  Oziernoy
1978 and Efstathiou and Silk 1983) accounts for the spin angular
momentum as a  remnant  of the primordial whirl and predicts that
the rotationalaxes of galaxies would be perpendicular to the main
plane of a large-scale structure.
 
    Another approach is represented by Zeldovich's pancake  model
(Sunayev  and Zeldovich 1972, Doroshkevich 1973, Shandarin  1974,
Doroshkevich,  Saar, and Shandarin 1978, Zeldovich  1978),  which
provides  that  structures  in the  Universe  form  "from  up  to
bottom".  In  the  effect of asymmetrical  collapse  of  a  large
structure  there arises a magnetohydrodynamic shock  wave,  which
causes the structure to fragment as well as imparts galaxies with
spin  angular momentum. In this scenario the mechanism of gaining
angular momentum by galaxies consists in acquiring rotation by  a
shock wave passing across the protostructure, which gives rise to
a  coherent,  non-random orientation of galaxy planes  in  space.
Thus  the  model  predicts the galaxies' rotational  axes  to  be
parallel to the main plane of the structure.
 
    It  should  be stressed that this classic view in which  each
of   the  basic  scenarios  predicts  a  different  alignment  of
rotational axes of galaxies was later somewhat disturbed. It  was
shown  that  in  principle  in any scenario,  including  that  of
hierarchical clustering, can happen a phase of shock wave,  which
may  give  rise  to alignment of rotational axes. Naturally,  the
extent of this alignment is differentiated.
 
    Two  main  methods of investigating alignment of galaxy  axes
have  been  proposed  to  date.  The  first  one,  consisting  in
examining  distributions  of position  angles  of  galaxies,  was
presented  in  detail by Hawley and Peebles  (1975).  Hawley  and
Peebles   (1975),   and  subsequently  Kindl   (1987)   discussed
detailedly  the  statistical methods they suggested  for  probing
position  angle  distributions. Unfortunately, the  distributions
provide  reliable data on galaxy plane orientation only  for  the
galaxies  observed  edge-on. In consequence,  the  galaxies  seen
face-on  or nearly face-on have to be excluded from the analysis.
Furthermore, in studying nearby structures the galaxies  have  to
be  located  in the vicinity of the structure's main  plane.  For
instance,  when  considering  galaxies  belonging  to  the  Local
Supercluster  (LSC),  we should take into account  only  galaxies
located at small supergalactic latitudes B  (God{\l}owski 1994).
 
    Jaaniste  and  Saar (1977, 1978) improved on a  novel  method
proposed by Oepik (1970), which took into account not just galaxy
position   angles  but  also  their  ellipticities   (angles   of
inclination to the line of sight). In this method, using galaxies
of  any  possible  orientation  and  location  on  the  sky,  one
calculates two angles. One is the angle between the normal to the
galaxy's plane and the main plane of the structure (polar  angle)
and  the other one is the angle between the normal projected onto
the  main  plane of the structure and the selected  direction  on
this plane (azimuth angle). For example, in the study of galaxies
belonging to the Local Supercluster it can be the line connecting
our   Galaxy  with  the  centre  of  the  Virgo  cluster.   After
determining  those angles, they can be analysed using statistical
methods to look for any non-random trends. Unfortunately, in this
method  for  any  galaxy  there are  two  possible  orientations,
corresponding  to  two possible positions of the  normal  to  the
galaxy  plane  in  space. Since usually  the  sense  of  galactic
angular  momentum is not known, there are actually four  possible
orientations  of a galaxy's rotation axis. Flin and  God{\l}owski
(1986)  improved  on  the  method of Jaaniste  and  Sarr  (1978),
correcting  the  errors  present  in  their  original  work,  and
described the method of analysing distribution of vectors  normal
to  the  galaxies'  planes,  basing on  the  statistical  methods
proposed by Hawley and Peebles (1975) and adapting them  for  the
purpose  of investigating angle distributions in this  case.  The
methods  were  further  developed  in  the  subsequent  works  by
God{\l}owski  (1993, 1994). In Jaaniste and Sarr's (1978)  method
one  has  to take into consideration both the fact that  galaxies
are flattened spheroids (Holmberg 1946), with their actual ratios
of   axes   depending  on  their  morphological  type  (Heidmann,
Heidmann, de Vaucouleurs 1971), and the Holmberg effect (Holmberg
1946,  1958, 1975, Fouque and Paturel 1985). The Holmberg  effect
consists in the fact that micrometric measurements of larger  and
smaller semiaxes of galaxies depend on ellipticity of the  image,
which  in  turn  creates a necessity of reducing the  micrometric
axis  ratio  q  to  the standard photometric system.  Fouque  and
Paturel (1985) derived the formulae for reduction of galaxy sizes
obtained   from   micrometric  measurements  in  various   galaxy
catalogues to a uniform system of photometric diameters. How  the
actual  ellipticity and the Holmberg effect influence the results
of   galaxy  orientation  analysis  was  discussed  in  Flin  and
God{\l}owski (1986, 1989a) and God{\l}owski (1993, 1994).
 
     The methods of investigating of galaxy orientation have been
further  scrutinized and improved ever since. A critical analysis
of  the  method of investigating spatial orientation of  galaxies
using   their  ellipticity  was  provided  in  God{\l}owski   and
Ostrowski (1999), where it shown that the process of deprojection
of  galaxies  basing on the catalogue data generates  significant
systematic  errors,  which  have to  be  taken  into  account  in
analysing   alignment  of  galaxy  rotation  axes.  The   effect,
confirmed  by  God{\l}owski, Baier and MacGillivray  (1998),  and
Baier,  God{\l}owski and MacGillivray (2003), could have  led  to
partly   distorting   the  earlier  results   concerning   galaxy
orientations.
 
    Cabanela   and   Dickey  (1999)  indicated  difficulties   in
detecting  rotational axis alignment for small samples. Analysing
a sample of 54 edge-on galaxies with spin vectors well-determined
from  HI observations, they demonstrated that even a considerable
amount  of anisotropy in distribution of galaxy spins within  the
Perseus  cluster, signalled in the work of Cabanel  and  Aldering
(1998), could have well remained undetected.
 
    Several variants of the methods mentioned above were used  in
investigating  rotational axis alignment by a number  of  authors
(see  e.g. Hu (1995, 1998), Wu (1997, 2006) Aryal Saurer (2000)).
Modifications of the original methods were also made in order  to
adapt  them  to some particular research problems  (Lee  and  Pen
2001,  Brown  et  al. 2002, Noh and Lee 2006a,  2006b,  Trujillo,
Carretero and Patiri 2006). And thus Lee and Pen (2001) described
their  method  of  investigating intrinsic  galaxy  alignment  in
respect  to the local tidal shear tensor and discussed in  detail
the  method  of  determining this tensor  as  well  as  examining
correlation both between the galaxy spins and between  the  spins
and  the  local tidal shear tensor. In turn, Brown et al.  (2002)
discussed  a  method  of  investigating  intrinsic  alignment  of
ellipticities  of  galaxies based on their ellipticity  variance.
Noh and Lee (2006a, 2006b) examined alignment of planes of spiral
galaxies seen edge-on in respect to the plane of local "pancake".
This  plane  was  determined locally for each individual  galaxy,
defined as a plane passing through the galaxy and its two nearest
neighbours.  Trujillo,  Carretero and Patri  (2006)  studied  the
distribution  of angles between rotational axes of  galaxies  and
the  vector connecting a given galaxy with the centre of a  local
"void"  (thus introducing a vector normal to the local  plane  of
the structure as defined by the ambient matter).
 
\section{Results of investigations of galaxy orientations in the Local Supercluster and its close neighbourhood.}
 
     The  orientation  of galaxies in the Local Supercluster  and
its close neighbourhood was studied by many researchers. Most  of
the  early  works on the alignment of rotational axes (Wyatt  and
Brown  1955,  Brown  1964, 1968, Reinhardt  1970,  Reinhardt  and
Roberts  1972, Nilson 1974, Jaaniste and Saar 1978,  MacGillivray
et  al.  1982a,b,  MacGilivray and Dodd 1985a,b,  Kapranidis  and
Sullivan 1983) implied that the galaxies are either oriented in a
random   way  or  the  galaxies'  planes  are  parallel  to   the
structure's  main plane. And thus MacGillivray et  al.  (1982a,b)
scrutinizing  the  distribution of spiral and irregular  galaxies
belonging  to  the  Local Supercluster found a  weak  correlation
between  the  galaxy  planes and the main  plane  of  LSC.  These
results  confirmed either the theory of hierarchic clustering  or
the  theory of primordial turbulence. At that time just  Jaaniste
and  Saar (1978), using not only position angles of galaxies  but
also their inclination, claimed that in LSC there is an excess of
galaxies with rotational axes aligned in the equatorial plane  of
LSC.
 
     The  question of existing discrepancies between the  results
of  early  research on galaxy orientation was  discussed  in  the
articles of Flin and God{\l}owski (1986) and God{\l}owski  (1993,
1994).  They  were caused mainly by contamination of the  samples
with  background  objects  as well as  difficulties  with  proper
interpretation  of  the  obtained  results  due  to   using   the
equatorial  coordinate system instead of the  supergalactic  one.
For instance, Kapranidis and Sullivan (1983) analysed the density
of  positions  of galactic poles within regions of  $30^o  \times
30^o$  size  for  a sample of spiral galaxies in  the  equatorial
coordinate  system. The authors claimed that there  had  been  no
departure  of  galactic  pole  distributions  from  the  expected
isotropic distribution, as they found an excess of galactic poles
in  only  two  of  the analysed regions, while it  escaped  their
notice  that these very regions ($\alpha = 195^o, \delta =  15^o$
and  ($\alpha = 15^o, \delta  = -15^o $) are in the direction  to
the  centre and anticentre of the Virgo cluster (i.e. the  centre
of  the  Local  Supercluster) respectively, which means  that  in
fact   they   discovered  that  the  galaxy  planes  prefer   the
orientation   perpendicular   to   the   plane   of   the   Local
Supercluster.
 
     The  question of proper choice of a coordinate system should
be  kept  in  mind  when  analysing orientation  of  galaxies  in
clusters  and superclusters other than LSC. In such  a  case  the
most appropriate one could be the coordinate system connected  to
the main plane of the analysed structure (Flin 1994, God{\l}owski
1995). The importance of proper choice of a coordinate system was
also  stressed  by  Bukhari  and  Cram  (2003),  Wu  (2006),  and
particularly  Aryal, Kandel and Saurer (2006),  who  showed  that
that   the  interpretation  of  the  results  of  probing  galaxy
orientation  in the Abell 3558 cluster - the core of the  Shapley
Concentration - depends on the choice of coordinate system.
 
    Flin  and  God{\l}owski (1986) and God{\l}owski (1993,  1994)
demonstrated  that  the planes of galaxies  belonging  the  Local
Supercluster  are oriented perpendicularly to the main  plane  of
the  LSC  equator,  while  the effect depended  on  the  galaxy's
morphological type. This result was later confirmed by the  study
of  Parnovski  et  al. (1994), which found an  excess  of  galaxy
rotational  axes directed towards $4-6^h, 20-40^o$.  The  authors
recognized  that the observed anisotropy is of global  character.
This  is generally consistent with the earlier results of  Fliche
and Soriau (1990) concerning the orientation of extended galactic
HI  envelopes and the "cosmic pole" detected in the  analysis  of
remote  quasars  ($5^h  30^m, \, 7^o$). However,  Flin  (1995)  upon
analysis  of these results indicated that the observed anisotropy
is  consistent  with the results of Flin and God{\l}owski  (1986)
and is not of global character but specific to LSC.
 
      However  these results seem to favour the pancake model  of
Zeldovich,  there are arguments that the situation is  much  more
complicated. Flin and God{\l}owski (1990) and God{\l}owski (1993)
studied   the  orientation  of  galaxies  belonging  to   various
substructures  of the Local Supercluster, defined basing  on  the
work of Tully (1986) and showed that the orientation is different
for  different  substructures. God{\l}owski (1994)  provides  the
dependence  of  orientation on radial velocity  of  the  galaxies
under   investigation.  Furthermore,  in  Flin  and  God{\l}owski
(1989a), Kashikawa and Okamura (1992) and  in God{\l}owski (1994)
it was  shown  that  the  rotational  axes  of  galaxies  at  low
supergalactic  latitudes are parallel to the main plane  of  LSC,
while  the  rotational  axes of galaxies  at  high  supergalactic
latitudes  are  perpendicular to that plane.  In  this  case  the
galaxy rotational axes tend to be directed towards the centre  of
the  Virgo cluster. This effect, confirmed recently by Hu et  al.
(2006), supports a more complicated hybrid model rather than  the
simple  pancake  model.  Also the results  of  Aryal  and  Saurer
(2005a), who found a weak preference of spin directions  for  the
spiral galaxies belonging to the Local Supercluster, do not agree
with the predictions of Zeldovich's pancake theory. Aryal and his
collaborators in the series of articles (Aryal, Kafle and  Saurer
2008,  Aryal, Neupane and Saurer 2008, Aryal, Paudel  and  Saurer
2008, Aryal 2010) analysed various samples of spiral galaxies  in
the   Local  Supercluster  and  its  neighbourhood  (with  radial
velocities   $V  <  5000 km/s$) and found that  only  the  barred
spirals  manifest  alignment of orientation of  galactic  planes,
while the alignment effect depend on their radial  velocities.
 
    On  the  other  hand, Lambas indicated that the alignment  of
galaxy  orientation can be of local character. Lambas, Groth  and
Peebles   (1988)  compared  the  position  angles  of  elliptical
galaxies from the UGC catalogue (Nilson 1973) in respect  to  the
neighbouring  large-scale structures  and  found  that  they  are
aligned  in the scale at least up to $2^o$, which corresponds  to
the  linear  scale $\sim 2 \times h^{-1} \, Mpc$.  Next  Muriel  and
Lambas  (1992),  taking  into account the  $3D$  distribution  of
galaxies  obtained  from  their redshifts,  determined  that  the
spiral  galaxies  are oriented towards their nearest  neighbours,
while  the  elliptical galaxies are aligned on the scale  smaller
than $3 h^{-1} \, Mpc$. Also Cabanela and Aldering (1998) noted that
the  anisotropy  they  found  in  the  distribution  of  galaxies
belonging  to the Perseus Supercluster can be of local,  and  not
global (for the entire supercluster) character.
 
    The  basic  mechanism  to  generate galaxy  rotation  in  the
scenario  of hierarchical clustering are the mutual tidal  torque
interactions  (Wesson (1982) and White (1984)  -  basing  on  the
ideas of Hoyle (1951)). In this scenario the distribution of spin
angular  momenta of galaxies should be random, but it  was  shown
that the local tidal shear tensor can cause a local alignment  of
rotational axes (Dubinski 1992, Catelan and Theuns 1996, Lee  and
Pen  2000, 2001, 2002, Navarro, Abadi and Steinmetz 2004).  Along
these  lines  - as consistent with the tidal torque  mechanism  -
were  also  reinterpreted  the earlier results  to  indicate  the
alignment of galaxies (e.g. those of Flin and God{\l}owski 1986).
On the other hand, some researchers, as e.g. Brook et al. (2010),
still  advocate the misalignment of angular momenta  that  occurs
during hierarchical structure formation.
 
    Brown  et  al.  (2002) studying ellipticity of galaxies  from
the  Supercosmos survey (mean redshift $z=0.1$) found  that  they
are  correlated  in  the scale between 1 and  100  arc  min.  The
meaning  of  this observation is not clear, however  the  authors
interpret  it  as a local effect, consistent with the  theory  of
tidal  interactions.  Lee  and Pen (2002)  attempted  to  measure
intrinsic alignment of galaxies out of observational data by  the
way  of  reconstructing the tidal shear tensor. However they  did
find  the  spin-shear correlation, their results are of  doubtful
value,  due to a high noise level (Lee 2004). Thus the  intrinsic
correlations between galaxy spins and intermediate principal axes
of  the tidal shears were confirmed only recently Lee and Erdogdu
(2007)  and  Lee  (2010). Navarro, Abadi  and  Steinmetz  (2004),
analysing  a  sample of near spiral galaxies seen edge-on,  found
that their planes are perpendicular to the main plane of LSC. Lee
(2004),  who interpreted this fact as consistent with  the  tidal
interactions   model,   basing  on  this  scenario   investigated
analytically its consequences for the large-scale orientation  of
galaxies   and  found  that  his  predictions  agree   with   the
distribution of orientations of nearby spiral galaxies. Trujillo,
Carretero  and Patiri (2006) determined, by probing the distribution
of angles between  rotational axes of galaxies and the vector
connecting a given galaxy with the centre of a local void, that the
spins  of  spiral galaxies located on the shells of  the  largest
cosmic  voids,  are  perpendicular to the direction  towards  its
centre  (i.e. along the direction of the envelope defined by  the
surrounding matter). Noh and Lee (2006a, 2006b), analysing nearby
galaxies from Tully's catalogue (Tully 2000), found alignment  of
spiral  galaxy  planes  perpendicular to the  plane  of  a  local
pancake,  which  they interpret along the lines of  the  recently
proposed  scenario of broken hierarchy (Bower et  al.  2005).  In
this  scenario, instead of hierarchical clustering  on  all  mass
scales,  there  is  an anti-hierarchical clustering  on  a  small
scale, since gravitational tidal effects give rise to objects  of
Zeldovich's pancake type (Zeldovich 1970) rather than objects  of
spherically  collapsing halo. However, contrary  to  the  classic
pancake  scenario,  this effect occurs just locally  on  a  small
scale.
 
Hernandez et al. 2007 using sample of 11 597 SDSS galaxies,
present formulas to derive an estimate of the halo spin parameters
$\lambda$  for any real galaxy, in terms of common observational
parameters. For spirals they found a good correlation between 
empirical values of $\lambda$  and visually assigned Hubble types,
showing usefulness of this physical parameter as one of classification
tools.
 
    Recently,  there  have  been  some  attempts  to  investigate
galaxy  angular  momenta  on a large scale.  Paz,  Stasyszyn  and
Padilla  (2008)  analysing galaxies from the  Sloan  Digital  Sky
Survey  catalogue  found  that the  galaxy  angular  momenta  are
aligned  perpendicularly to the planes of large-scale structures,
while  there is no such effect for the low-mass structures.  They
interpret this as consistent with their simulations based on  the
mechanism  of  tidal interactions. Jones, van  der  Waygaert  and
Aragon-Calvo  (2010)  found that the  spins  of  spiral  galaxies
located within cosmic web filaments tend to be aligned along  the
larger  axis of the filament, which they interpreted as  "fossil"
evidence indicating that the action of large scale tidal  torques
effected the alignments of galaxies located in cosmic filaments.
 
The largest scale alignment was found in the series of paper by
Hutsemekers (Hutsemekers 1998, Hutsemekers and Lamy 2001,
Hutsemekers et al. 2005) during analyzis of the  alignment of
quasar polarization vectors.
In the paper Hutsemekers et al. (2005), based on a new sample of
355 quasars with significant optical polarization they fount that
quasar polarization vectors are not randomly oriented over the sky.
The polarization vectors appear coherently oriented or aligned over
huge ($\sim$ 1 Gpc) regions of the sky. Furthermore, the mean
polarization angle $\theta$ appears to rotate with redshift at the
rate of $\sim 30^o$ per Gpc. While interpretations like a global
rotation of the Universe can potentially explain the effect, the
properties they observed qualitatively correspond to the dichroism
and birefringence predicted by photon-pseudoscalar oscillation within
a magnetic field. These results usually are not questioned, (with
exeptions of  Joshi et al. 2007 which not found any effects during
analyzis of theirs sample (Jackson et al. 2007)), however the origin
of this effect is still discussed. Possible interpretations of the
alignment effect have been discussed in Hutsemekers (1998),
Hutsemekers and Lamy (2001), Hutsemekers et al. (2005) and more
recently by several authors (Jain et al. 2002, 2004, Ralston and Jain 2004,
Hutsemekers et al. 2010, Poltis and Stojkovic 2010, Silantev et al 2010,
Antoniou and Perivolaropoulos 2010, Agarwal, Kamal and Jain 2011).
 
     Recognizing  intrinsic alignment of  galaxy  spins  is  also
important  with  regard to investigations of  weak  gravitational
lensing,  in  which, according to Heavens, Refregier and  Heymans
(2000),  the intrinsic spin alignment in galaxy pairs has  to  be
taken  into account, since otherwise it would affect the  results
in  a  systematic way. Crittenden et al. (2001)  proved  that  at
least  in  the  scenario of tidal interactions,  the  effects  of
alignment  can  be  distinguished  from  the  effects   of   weak
gravitational  lensing, while Heymans et al.  (2004)  provided  a
detailed  analysis  of  some method to determine  intrinsic  spin
orientation  in galaxy pairs as well as to remove them  from  the
weak  gravitational lensing considerations.  From  our  point  of
view,  it  is of importance that they thus confirmed the presence
of  spin  alignment  in  pairs  of galaxies.  Systematic  effects
occurring in weak gravitational lensing investigations were  also
discussed by Mandelbaum et al. (2005).
 
    This   brief   review  of  results  of  studies   of   galaxy
orientations  indicates  that there  are  no  deciding  arguments
weighing in favour of any galaxy formation scenario to date. Most
of  them  can  be  interpreted as supporting either  the  pancake
scenario  or the scenario of tidal interactions. One should  note
the  fact that while in the pancake scenario involves an explicit
mechanism   to   generate  non-random  orientations   of   galaxy
rotational  axes, there is no obvious mechanism of this  kind  in
the  other  scenarios. Then it is necessary  to  more  accurately
recognize  galaxy orientations within the structures of  galaxies
in  order  to  look for further evidence supporting  one  of  the
scenarios.  Such  evidence  could be  provided  by  investigating
orientation of galaxies within clusters.
 
\section{Results  of  investigation of galaxy orientations in clusters of galaxies}
 
    The  investigations of galaxy orientations in  clusters  have
also  quite  a  long history. Thompson (1976) found alignment  of
galaxy orientations in the Virgo and A2197 clusters. Adams (1980)
discovered   a   bimodal  distribution  of  galaxy   orientations
examining the combined data for seven galaxy clusters (A76, A179,
A194,  A195,  A999, A1016, A2197). The orientation  of  principal
axes of the clusters corresponded with one of those maxima. Helou
and  Salpeter (1982), studying 20 galaxies belonging to the Virgo
cluster,  found  that  their spins are not  directed  in  random,
however  the nature of this non-random distribution was  not  too
clear.   Mac   Gillivray  and  Dodd  (1985a)   investigated   the
distribution of orientation of galaxies in the Virgo cluster  and
showed that the galaxy planes are perpendicular to "the direction
towards the cluster's centre", i.e. the galaxies' rotational axes
are  aligned  towards that centre. Gregory,  Thompson  and  Tifft
(1981)  found  in the Perseus supercluster (A426, A262)  and  the
groups   around  the  galaxies  NGC383  and  NGC507   a   bimodal
distribution  of  galaxy position angles  with  two  maxima.  The
position of one of them corresponds to the position angle of  the
Perseus  supercluster. Here we have two populations  of  galaxies
with almost perpendicular  axes.
 
    On  the  other hand, Bukhari (1998), as well as  Bukhari  and
Cram (2003) studying orientation of galaxies within clusters  did
not recognize any alignment. Han, Gould and Sackett (1995) probed
a  region  of  the  Local Supercluster with an  enhanced  density
galaxies containing 94 objects. Analyzing a sample of 60 galaxies
for which accurate values of spins were available, they found  no
alignment. Hoffman et al. (1989) for a sample of 85 galaxies with
well-known   spins  from  the Virgo  cluster  did  not  find  any
alignment  either. Flin and Olowin (1991), Trevese, Cirimele  and
Flin  (1992), Kim (2001), investigating isolated Abell  clusters,
detected  just  rudimentary  traces  of  alignment.  The  similar
results were obtained by Torlina, De Propris and West (2007) from
studies  of  the  Coma   cluster and its vicinity.  Gonzalez  and
Teodoro  (2010) interpreted the alignment of just  the  brightest
galaxies within a cluster as an effect of action of gravitational
tidal   forces.  Correlation  between  the  orientation  of   the
brightest  galaxy within a cluster and the cluster's  large  axis
was  also  found  by Sastry (1968), Carter and  Metcalfe  (1980),
Binggeli  (1982),  Struble and Peebles (1985), Rhee  and  Katgert
(1987),  West  (1989, 1994), van Kampen and Rhee (1990),  Plionis
(1994), Fuller et al. (1999) and Kim et al. (2002).
 
    God{\l}owski  and  Ostrowski  (1999)  analysed   in   various
coordinate  systems orientation of galaxies in 18 Tully's  groups
within  the  Local  Supercluster  taken  from  Tully's  catalogue
(1988).   As  the  fundamental  coordinate  system,  the  second,
supergalactic   coordinate   system   was   assumed   (Flin   and
God{\l}owski 1986) and the position of the coordinate system pole
was varied (by $5^o$ at a time), both in supergalactic latitude B
and  in  longitude  L. If the galaxy rotational  axes  preferably
aligned  in  some plane, there would be a deficit  of  rotational
axes  oriented  in  the direction perpendicular  to  that  plane,
whereas  if the orientations of galaxy rotational axes  preferred
some  particular  direction, then there should be  an  excess  of
galaxies  with  spins  oriented along this  direction.  A  strong
correlation was found between those maxima and the line of  sight
towards  the  cluster, while there were no such correlations  for
the  other  analysed directions. The obtained  excess  of  galaxy
rotational  axes  oriented along the line of  sight  towards  the
cluster  can  suggest that we are dealing here with a  systematic
effect related to the process of deprojection of axes of galaxies
from  their  optical images. The essential result of God{\l}owski
and  Ostrowski (1999) consists in discerning within the catalogue
data  a  strong  effect (further called the  line-of-sight  (LOS)
effect)  related to the process of deprojecting axes of  galaxies
from their optical images. It was shown that the very process  of
deprojection  of  galaxies  off  the  catalogue  data   generates
considerable  systematic errors, which  have  to  be  taken  into
account  in  investigating  galaxy  orientations.  The  discerned
effect  mostly  obscures  any possible  weak  effects  of  galaxy
orientation alignment in the examined clusters. The problem  with
deprojection  of  optical images had been  earlier  signalled  by
other  authors, particularly Kapranidis and Sullivan  (1983).  In
the work mentioned above (God{\l}owski and Ostrowski 1999) it was
proved  that  the  effect is significant  for  Tully's  catalogue
(1988). The main reason behind this is that in calculating galaxy
inclination angles Tully assumed that the "true" ratio of axes of
galaxies is 0.2, which is a rather poor approximation, especially
for non-spiral galaxies (God{\l}owski 2011).
 
    This  result presented above, pertains to the analysis basing
on Tully's catalogue (1988), showing that in the data there is an
excess  of galaxies seen face-on. It should be stressed  that  by
modelling the LOS effect for a sample galaxies taken off the  UGC
(Nilson  1973)  and ESO (Lauberts 1982) catalogues and  including
the effects of true ellipticity (after Heidmann, Heidmann and  de
Vaucouleurs  (1971))  and  Holmberg  (after  Fouque  and  Paturel
(1985)) it was found that the LOS effect, if any, is much smaller
(God{\l}owski  and Ostrowski 1999, God{\l}owski  2011).  This  is
consistent with the result of Bahcall, Guhathakurt and  Schneider
(1990),  who stated that there is no excess of elliptic  galaxies
seen  face-on in the UGC, where is actually an excess  of  spiral
galaxies seen edge-on, which can be accounted for either  by  the
Holmberg effect or by "dropping" weak spiral galaxies, classified
as  stars, off the catalogue. To summarize, the results show that
in examining galaxy orientations using their ellipticity, one has
to  make  a  very careful analysis of the available observational
data,  especially  in regard of the Holberg and true  ellipticity
effects,  as  well as of possible occurrence of other  systematic
observational effects.
 
       Problem of "true" shape of galaxies is also very important
in the context of Tully Fisher (Tully and Fisher 1977) and Baryonic
Tully Fisher relation (Freeman 1999, Mc Gaugh et al. 2000 and
Gurovich et al. 2010 for the latest review). One of important problems
for these relations is correction to edge-on inclination. The
inclination angles are obtained by assuming that the "true" ratio
of axes of galaxies is 0.2, which, as it was noted above, is
a rather poor approximation. The Tully Fisher relation is an
important point where observations and the theory of galaxy angular
momentum meet.
 
    Orientation of galaxies in Tully's groups was examined  again
in God{\l}owski, Szyd{\l}owski and Flin (2005). In conclusion, it
can  be  stated  that  no  significant  evidence  was  found   to
demonstrate  alignment of galaxy rotation axes  in  the  analysed
Tully's  groups. Also Aryal and Saurer (2005c), who  investigated
three  Abell  clusters of richness class zero S0794,  S0797,  and
S0805, found that the galaxies in those poor cluster are oriented
at random.
 
    God{\l}owski, Baier and MacGillivray (1998) inspected  galaxy
orientations  within the cluster Abell 754 basing on  COSMOS/UKST
Southern  Sky  Objects  Catalogue  (Yentis  et  al.  1991).  They
analysed the distributions of galaxy position angles as  well  as
polar and azimuth angles, confirming the LOS effect recognized by
God{\l}owski  and  Ostrowski (1999), which generates  significant
systematic  errors  in  the  process of  deprojection.  For  that
reason, the conclusions concerning galaxy orientations within the
clusters  under examination were based on the analysis of  galaxy
position  angles.  In  this work, it was  demonstrated  that  the
distribution  of galaxy orientations in the double cluster  Abell
754  is  non-random, with the galaxy planes perpendicular to  the
main  plane  of the cluster. The same method was applied  to  the
cluster  Abell  14  (Baier, God{\l}owski and MacGillivray  2003),
where  a non-random distribution of galaxy orientation was  found
too,  but  the  direction of this alignment (in  respect  to  the
cluster's  main plane) is  perpendicular to that  in  Abell  754.
According  to  Di  Fazio and Flin (1988), such a  discrepancy  in
preferred directions of galaxy orientation in respect to the main
plane  of  the  cluster can be related to existing two  types  of
galaxy clusters: oblate and prolate ones.
 
    While   examining   galaxy  orientation  in   the   clusters'
substructures,  it  was  found that for  Abel  754  and  Abel  14
clusters, the galaxies within their substructures are oriented in
a  non-random  way, consistent with the orientation  of  galaxies
within  the entire cluster. A higher-mode analysis of the Fourier
test  ($4 \theta$) shows that the position angle distribution  in
those   clusters  can  have  even  more  sophisticated,   bimodal
character,  probably  related to the presence  of  substructures.
However,  since  the maxima in the distribution  for  the  entire
cluster and its principal substructure are the same, this  latter
result should be regarded with caution.
 
    It  was  confirmed  that  there is an  effect  recognized  by
God{\l}owski  and Ostrowski (1999), giving rise  to  considerable
systematic errors in the process of galaxy deprojection from  the
catalogue  data. As this confirmation is based on the independent
observational data, it means that the effect occurs not only  for
the  data from Tully's catalogue (1988). Thus it could contribute
to  distortion  of some of the earlier results concerning  galaxy
orientation.  The  main reason that the effect  occurs  with  the
COSMOS/UKST Southern Sky Objects Catalogue (Yentis et  al.  1991)
data,  is apparently the fact that the catalogue does not provide
morphological types of galaxies, making us to assume for all  the
galaxies that the "true" galaxy axis ratio $q_0 = 0.2$, which is,
as  it  was  already  mentioned,  a  rather  poor  approximation,
especially  for  non-spiral  galaxies. However,  the  effect  was
found  as  much stronger than for Tully's groups in  God{\l}owski
and  Ostrowski (1999). It may be due to the fact that COSMOS  has
difficulties  with  distinguishing  between  stars  and  galaxies
(Bukhari  and Cram 2003). In effect, the sample under examination
can be contaminated by stars misclassified by COSMOS as galaxies.
Furthermore, Cabanela and Aldering (1998) showed that some effect
similar to the Holmberg effect can be considerable also when  the
data  are  taken  off  a  catalogue in which  galaxy  sizes  were
measured  through  the  method  of digital  scanning.  Using  the
Minnesota  Automap Plane Scanner (Pennington and al.  1993)  they
compiled   the   catalogue   of   galaxies   belonging   to   the
Pisces-Perseus  supercluster,  and  basing  on  the  analysis  by
Huizing (1994) noted that the external regions of spiral galaxies
(in  contrast to their internal regions) are optically thin,  and
thus  the  measured galaxy diameters should increase  with  their
inclination,  which  in  turn leads to  their  overrepresentation
after cataloguing. However Cabanel and Aldering's analysis (1998)
should  be  regarded with caution, it is evident the effect  they
found  has  to  be  taken  into  account  in  galaxy  orientation
investigations  based on their ellipticity,  when  the  data  are
taken  off  catalogues  in which galaxy  sizes  are  measured  by
digital scanning.
 
    Summing  up the results obtained by various authors,  it  can
be  stated  that we have no satisfactory evidence to support  the
galaxy  axis  alignment  in  the  groups  and  poor  clusters  of
galaxies, while there is ample evidence of this kind for the rich
clusters of galaxies. However, it should be emphasized that  even
where  actually  is  alignment,  there  is  no  common  preferred
direction  of  galaxy spin orientations, and thus while  in  some
clusters  the  direction perpendicular to the main plane  of  the
cluster  is preferred, in other cluster the galaxy spins tend  to
be  oriented parallelly to the structure's main plane. The  cited
above  results  are  not consistent with any of  the  three  main
scenarios of galaxy formation.

\section{Angular momenta of galaxies - observational aspects of Li's model}
 
    If  we  now  assume  validity of Li's model,  then  from  the
analysis of observational data on orientation of galaxies it  can
be  supposed that the structures with a mass typical  for  galaxy
groups  are  just  the structures with minimal angular  momentum.
For this reason, the orientation of galaxies in 18 Tully's groups
was  examined once again in God{\l}owski, Szyd{\l}owski and  Flin
(2005) with respect to distribution of angles: polar, azimuth and
position ones in the supergalactic coordinate system for galaxies
belonging  to  those  clusters,  using  three  statistical  tests
($\chi^2$, Fourier, and autocorrelation). The inspection of polar
and  azimuth  angle distributions indicate possible occurring  of
rotation  axis  alignment in some of the analysed Tully's  groups
(groups 11, 31, 41, 51, 52). One should however have in mind that
in  the  catalogue  data (Tully 1988) there is a  strong  effect,
recognized by God{\l}owski and Ostrowski (1999), related  to  the
process  of  deprojection of axes of galaxies from their  optical
images.  Therefore  the  distributions of supergalactic  position
angles were analysed independently. The results obtained from the
analysis  of  position angles do not confirm  the  hypothesis  of
galaxy plane orientation alignment. In effect, it can be asserted
that  no  sufficient  evidence  was  found  to  support  such  an
alignment  in  Tully's  groups under examination.  It  should  be
stressed that the most significant effect was discovered  in  the
most  massive  structure  -  Tully's  group  no.  11  (the  Virgo
cluster).
 
    In  the  cited  work  (God{\l}owski, Szyd{\l}owski  and  Flin
2005),  the  authors  considered also the  relation  between  the
angular  momentum and the mass of a structure  as  based  on  the
observational data. It was investigated how this image changes in
dependence  on  the  mass  of structures  made  up  of  galaxies,
beginning  from the simplest ones, i.e. pairs of galaxies.  These
investigations  showed  that  their angular  momentum  originates
mainly  from  the  orbital motion of galaxies  (Karachentsev  and
Mineva  1984a,  1984b, Mineva 1987). Helou  (1984),  examining  a
sample  of  31  galaxy  pairs, found that an "anti-alignment"  of
spins  of  these  galaxies. Parnovsky,  Kudrya  and  Karachentsev
(1997) recognized a weak alignment in physical pairs of galaxies.
Alignment  in  pairs  of spiral galaxies was  also  discerned  by
Pestana and Cabrera (2004). Intrinsic spin alignment in pairs  of
galaxies  was  independently confirmed by Heymans et  al.  (2004)
within their research on weak gravitational lensing, where it was
necessary to estimate and remove the effects related to alignment
of  galaxy  orientations. Also the analysis of positions  of  the
Milky  Way's companions shows their non-random distribution (they
are  located perpendicularly to the Milky Way's disc), which  can
be  regarded as their orbital alignment. Galaxies within  compact
groups   rotate  on  prolate  orbits  about  the  group's  centre
(Tovmassian et al. 2001), which contributes to the system's total
angular  momentum. Thus it can be maintained that  structures  as
galaxies  and  their  companions, pairs of galaxies  and  compact
groups of galaxies have a non-zero net angular momentum.
 
    The  more  massive  structures are  groups  and  clusters  of
galaxies,  for which there is no evidence of rotation.  Moreover,
Hwang  and Lee (2007) examined dispersions and velocity  gradient
of  899 Abell clusters and found a possible rotation in only 6 of
them.  Any  non-zero angular momentum of groups and  clusters  of
galaxies would just come from possible alignment of galaxy spins.
There is no sufficient evidence to support galaxy rotational axis
alignment  in groups and poor clusters of galaxies. Additionally,
it  is  obvious  that  in  the isolated  Abell  groups  just  the
brightest  galaxies  manifest a rudimentary alignment  (Flin  and
Olowin 1991, Trevese, Cirimele and Flin 1992, Kim 2001), while in
the  most  numerous  clusters as A754  (God{\l}owski,  Baier  and
MacGilivray  1998),  A14  (Baier,  God{\l}owski  and  MacGilivray
2003),  A1656  (Djorgovski 1983, Wu et al. 1998, Kitzbichler  and
Saurer  2003),  a  non-random galaxy  orientation  alignment  was
found.  The most recent analysis of alignment in the Coma cluster
was  carried  out  by  Adami et al. (2009),  who  concluded  that
alignment  occurs  only in some regions of the  cluster  and  the
galaxies   of  early  and  late  spectral  types  have  different
orientations. Similarly, the preferred orientation  is  different
in  various parts of the cluster. The alignment of galaxy  planes
was  recently found in the cluster 1689 (Hung and in 2010).  This
result is of importance as it has been the most remote cluster to
have  been  looked for the alignment effect to date. Presence  of
non-random galaxy spin orientation has been ascertained  both  in
the  Local Supercluster and other superclusters, as the  Hercules
supercluster  (Flin 1994), Coma/A 1367 (Djorgovski 1983,  Garrido
et  al.  1993,  Wu  et  al.  1997, Flin  2001)  and  the  Perseus
supercluster (Gregory, Thompson and Tift 1981, Flin  1988,  1989,
Flin and God{\l}owski 1989b, Cabanela and Aldering 1998).
 
    In  God{\l}owski, Szyd{\l}owski and Flin (2005) the following
observational situation was depicted. For such galaxy  structures
as  pairs  of  galaxies and their compact groups, there  is  some
evidence  to indicate a non-zero angular momentum of  the  entire
structure, while for the more massive structures, groups and poor
clusters  of  galaxies  evidence of this  kind  is  absent.  From
examining rich galaxy clusters and superclusters the results were
obtained,  which  showed  a non-random  orientation  of  galaxies
within   these  structures,  imparting  a  non-zero  net  angular
momentum.  Such dependence of a structure's angular  momentum  on
its mass is consistent with the conclusions from Li's model.
 
    This  overall image was supported in subsequent studies  into
the  topic  of  galaxy orientations. For instance,  Yang  et  al.
(2006)  found,  while Sales and Lambas (2009)  and  Wang  et  al.
(2009,  2010)  confirmed it, that the companions of  central  red
galaxies are aligned along their large axes. Aryal and Saurer  in
a  series of articles (Aryal and Saurer 2004, 2005b, 2005c, 2006,
Aryal,  Pudel and Saurer 2007) dealt with orientation of galaxies
in  32 clusters of various numerousness, confirming (Aryal, Pudel
and  Saurer  2007) the situation that we do not find  any  galaxy
orientation alignment in sparsely populated clusters, while  such
alignment  is observed in a number of rich clusters of  galaxies.
However, either in God{\l}owski, Szyd{\l}owski and Flin (2005) or
in  Aryal,  Pudel  and Saurer (2007) only a qualitative  and  not
quantitative analysis on this issue was provided.
 
    Therefore God{\l}owski et al. (2010) examined orientation  of
galaxies  in  clusters both qualitatively and quantitatively  and
found  a  sharp  increase  of galaxy orientation  alignment  with
numerousness of the cluster. The analysis was carried  out  on  a
sample  of  247  Abell clusters with at least 100  objects  each,
taken  off  the Panko and Flin catalogue (2006), which  had  been
based on the Muenster Red Sky catalogue (MRSS), covering the area
of  5000 square degrees on the southern sky and providing data on
5.5 million galaxies. The catalogue is statistically complete  to
the  value of  $m=18.3$ mag. The PF catalogue includes structures
with  at least 10 members with luminosities within the range from
m3  through $m_3+3$, where m3 is the value of luminosity  of  the
third  brightest galaxy within a given region. The clusters  were
selected  using  the  tessellation  method  of  Voronoi   (Panko,
Juszczyk and  Flin 2009).
 
    For  each cluster analysed in God{\l}owski et al. (2010), the
distributions of position angles as well as of polar and  azimuth
angles  were  determined,  both in  the  equatorial  and  in  the
supergalactic  coordinate  system. In  analysis  of  these  angle
distributions the statistics of $\chi^2$ and Fourier  tests  were
applied  (Halley  and Peebles 1975, Flin and  God{\l}owski  1986,
God{\l}owski 1993, 1994). The values of analysed statistics  were
found  to  increase  sharply with the  numerousness  of  a  given
cluster, while this effect was stronger for the polar and azimuth
angles  than  for  the  position ones, and still  stronger  after
limiting  to  galaxies more luminous than $m_3+3$,  which  proves
that  the  effect  is  actually related  to  the  clusters  under
examination.
 
    In  God{\l}owski et al. (2010) it was found that, contrary to
the  suggestions of Aryal and Saurer (2004, 2005b, 2005c,  2006),
Aryal,  Pudel  and  Saurer (2007), the orientation  alignment  of
galaxies  is  weakly  correlated with their  morphological  types
according  to the classification of Bautz-Morgan (BM), while  the
suggested by Plionis et al. (2003) correlation between the degree
of  galaxy  orientation alignment and the  dispersion  of  galaxy
velocities in clusters was not confirmed.
 
    However  the results of Aryal and Saurer (2006)  as  well  as
those   of   God{\l}owski  et  al.  (2010)  are  consistent   the
predictions of  Li's model, they may be as well interpreted as an
effect of tidal forces, according to the scenario of Catelan  and
Theuns  (1996). The analyses of Noh and Lee (2006a,  2006b)  also
suggest that the linear theory of tidal interactions there should
involve  a relation between the galaxy orientation alignment  and
the structures' masses.
 
     It  should  be  emphasized  that  Li's  model  is  obviously
disputable  and cannot provide explanation for all  the  observed
effects,  but  it  is still of interest as it  accounts  for  the
dependence of structures' angular momentum on their masses, which
the other models have been not able to explain so far.
However please note that the simple relation between the mass of
a cosmic  structure and its angular  momentum  expressed by the
empirical relation $J\sim  M^{5/3}$  is also predicted by Tidal
Torque scenario.

\section{Summary}
 
     One  of the key issues of modern extragalactic astronomy  is
the  problem of origin of galaxies. This work describes in detail
various  hypotheses  of  galaxy  formation  and  discusses  their
connection to galaxy orientation. There are several scenarios  of
galaxy  formation,  which  can explain a  number  of  theoretical
problems  and observational results. However, none of  them  does
account  for all the difficulties concerning the process to  form
galaxies  and their structures. The purpose of this  work  is  to
present  the  current observational status of galaxy orientation,
and  subsequently to check to what extent the existing models  of
formation  of  galaxies and their structures are consistent  with
observations. It also describes the theoretical and observational
aspects  of  dependence of the structures'  angular  momentum  on
their   mass   and  comprehensively  discusses  the  results   of
investigations  of galaxy orientation within various  structures,
finding that Li's theoretical model, in which galaxies form in  a
rotating universe, is supported by the observational evidence. An
overview   of  possibilities  and  theoretical  difficulties   in
constructing cosmological models involving rotation as well as of
observational constraints on the actual amount of rotation of the
Universe is also provided.
 
      Studying  orientation  of  galaxies  is  one  of  the  most
effective  methods of testing the galaxy formation scenarios.  In
this  paper  the  observational results of the investigations  of
angular momentum of galaxies was discussed and compared with  the
predictions  of  three classical scenarios of  galaxy  formations
i.e.  primordial  turbulences,  Zeldovich's  pancake  model   and
hierarchical  clustering  with mutual tidal  torque  interactions
mechanism.   The results of early investigations of alignment  of
rotational axes within cosmic structures were usually interpreted
as  consistent  with Zeldovich's pancake model or  the  model  of
tidal   torque   interactions  in  the  scenario  of   hierarchic
clustering.  Recently,  some  variants  of  tidal  torque  model,
usually  with  hierarchical  clustering  scenario,  have   become
commonly  accepted. Only the pancake model involved some  obvious
mechanism  of acquiring angular momentum by galaxies. Admittedly,
in  the  most recent model of broken hierarchy the tidal  effects
also give rise to object of the kind of Zeldovich's pancakes (but
on  a  small scale), however all the main models of formation  of
galaxies   and  their  structures  in  the  Universe  encountered
difficulties in accounting for the  observed dependence of the
structures' angular  momentum  on  their masses.
Tidal Torque scenario predict that the relation between the
mass of a cosmic  structure and its angular momentum is expressed
by the empirical relation $J\sim  M^{5/3}$.
In Li's  model, in which
galaxies  form  in a rotating universe, there is a quite  evident
mechanism to impart the galaxies with angular momentum. Moreover,
it  provides for an attempt to relate the angular momentum  of  a
structure to its mass. If only for these reasons, the model is of
interest,  while naturally it has its weak points. Its  essential
drawback  consists  in  the  fact that  the  observed  amount  of
rotation  of  spiral  galaxies cannot arise from  the  Universe's
rotation  alone,  since the required amount of  rotation  of  the
Universe   is   too  large  in  comparison  with   the   detected
anisotropies of cosmic background radiation.
 
     The  work  reveals  that the presence  of  non-zero  angular
momentum  is contingent upon the structure's size. The  presented
observational  situation is that the galaxies,  their  pairs  and
compact  groups  have a non-vanishing angular  momentum.  In  the
structures  of  mass  corresponding to groups  of  galaxies  this
feature   is  not  to  be  found,  while  in  the  clusters   and
superclusters alignment of galaxy orientation has been discerned,
which  indicates a non-zero net angular momentum.  It  was  shown
that  such  an observational depiction of dependence  of  angular
momentum on the structure's size is consistent with Li's model.
 
    Li's  model  assumes  the global rotation  of  the  Universe.
Unfortunately,  there is no relatively simple model  of  rotating
universe  derived  from the general relativity theory,  involving
observables  that  could  be  used  for  its  empirical  testing.
However,  we  can make some estimates of global rotation  of  the
Universe  on the grounds of the Newtonian equivalent  of  general
relativity, while being fully aware that the Newtonian  cosmology
does  not  allow for describing actual evolution of the Universe,
being   but   its  conveniently  applicable  approximation.   The
situation   is   somewhat   complicated   by   the   fact    that
observationally  it is very hard to distinguish  the  effects  of
global   rotation  from other effects, as the Casimir  effect  or
dark radiation in the brane scenario of Randall-Sundrum. One  can
still  estimate from the observational data the net  contribution
of  all  effects  of this kind. It is shown  that  if  it  has  a
non-zero  value,  then one can explain the discrepancies  between
the  values  of matter density $\Omega_{m,0}$ and of  the  Hubble
constant  derived from measurements of SNIa and CMBR.  It  allows
also  to  account for the discrepancies between the observed  and
theoretical abundances of helium-4 and deuterium.
 
The main assets of this work are:
\begin{enumerate}
\item  It  was  recognized that the presence of a negative  term
scaling as $(1+z)^4$ in the Friedmann equation (or its Newtonian
equivalent) can be of various origin. Among the possible causes
could be the cosmological model with global rotation, the FRW
model in the brane scenario of  Randall-Sundrum with dark
radiation and the Casimir effect. Due to this, it is virtually
impossible to separate the effects of rotation using the standard
astronomical tests.
\item   Using the observational data concerning the Ia-type
supernovae, radio galaxies, cosmic background radiation and
baryon oscillations, a constraint on the net contribution of
density parameters of fluids of negative density scaling as
$(1+z)^4$ was determined.
\item   It was shown that the presence of such fluids of negative
density allows for obtaining a flat model of the Universe with
$\Omega_{m,0}=0.3$, consistent both with the Ia-type supernovae
and CMBR observations.
\item   It was demonstrated that analysing the first peak in the
CMBR power  spectrum one gets a much stricter constraint on the
sum of density parameters of negative fluids scaling as radiation
than one derived earlier from the nucleosynthesis analysis.
\item   Observational results are presented to indicate that the
galaxy orientation alignment is contingent upon the mass of a
given structure: for Tully's groups of galaxies no satisfactory
evidence was found for galaxy plane alignment, while there are
non-random orientations of planes of galaxies belonging to the
Abell clusters.
\item   The observational situation was depicted as follows: pairs
and compact groups of galaxies reveal a non-vanishing angular
momentum; for groups of galaxies there is no evidence for a
non-zero angular momentum, while such evidence can be provided
for clusters and superclusters of galaxies.
\item   It was shown that the galaxy orientation alignment increases
with the numerousness of clusters, while it is weakly correlated
with their morphological  type according to the Baitz-Morgan (BM)
classification.
\item   It was shown that Li's model, in which galaxies form in a
rotating universe, predicts dependence of angular momentum on the
structures' mass, which is consistent with the observational
circumstances mentioned above.
\item   The deprojection effect was discussed, which can essentially
modify the results of some earlier studies concerning galaxy
orientation.
 
\end{enumerate}
 
   It  should  be explicitly stressed that none of the  available
scenarios  of formation of galaxies and their structures  in  the
Universe  provides a complete explanation for  all  the  observed
properties of the object under consideration. In particular, this
is  the  case  with  observations of  galaxy  orientation  within
structures  of  galaxies, while various models  can  account  for
various  aspects  of  the galaxy formation process  on  different
scales,  as  well  as  for  various  observational  features   of
structures.

\section*{Acknowledgments}
 
The  author thanks prof. Piotr Flin for discussion.
The author thanks the anonymous referee for suggestions and  comments
improving the paper.

\end{document}